# Modelling China's Credit System with Complex Network Theory for Systematic Credit Risk Control

LU Xuan[3], HUANG Li[1], LYU Kangjuan[1, 2],

(1. The School of Economics, Shanghai University, Shanghai, 200444;
2. SILC Business School, Shanghai University, Shanghai, 201800;
3. Independent Researcher, Sydney, 2000)

**Abstract**

The insufficient understanding of the credit network structure was recognized as a key factor for regulators' underestimation of the destructive systematic risk during the financial crisis that started in 2007. The existing credit network research either took a macro perspective to clarify the topological properties of financial systems at a descriptive level, or analyzed the risk transmission path and characteristics of individual entities with much pre-assumptions of the network. Here, we used the theory of complex network to model China's credit system from 2000 to 2014 based on actual financial data. A bipartite financial institution-firm network and its projected sub-networks were constructed for an integrated analysis from both macro and micro perspectives, and the relationship between typological properties and systematic credit risk control was also explored. The typological analysis of the networks suggested that the financial institutions and firms were highly but asymmetrically connected, and the credit network structure made local idiosyncratic shocks possible to proliferate through the whole economy. In addition, the Chinese credit market was still dominated by state-owned financial institutions with firms competing fiercely for financial resources in the past fifteen years. Furthermore, the credit risk score (CRS) was introduced by simulation to identify the systematically important vertices in terms of systematic risk control. The results indicated that the vertices with more access to the credit market or less likelihood to be a bridge in the network were the ones with higher systematically importance. The empirical results from this study would provide specific policy suggestions to financial regulators on supervisory approaches and optimizing allocation of regulatory resources to enhance the robustness of credit systems in China and in other countries.

1. Introduction

In the financial crisis that started in 2007, the defaults and losses of some individual banks were magnified rapidly through the credit network among financial institutions to the entire financial marketing, leading to a dramatic impact to the whole economy. The after-event analysis regarded the insufficient understanding of the credit network structure as a key factor for regulators' underestimation of the destructive systematic risk (Brunnermeier, 2008). In response to the crisis, the FSB (Financial Stability Board) together with BCBS (Basel Committee on Banking Supervision) have published a series of important documents to manage systematic risks, such as *Global Systemically Important Banks: Updated Assessment Methodology and the Higher Loss Absorbency Requirement*. In China, the *China Financial System Stability Assessment* (2012) compiled by World Bank and the People's Bank of China suggested that the Chinese financial system has become more vulnerable to systemic risk than before. For effective control of the systematic risk, while the correlations among financial institutions have become increasingly strengthened and complicated, the understanding of the financial networks' typological properties and the risk transmission path would be of great use for policy suggestion.

The network analysis has been increasingly used in finance area to research financial networks and their dynamic evolution procedure in recent years. Integrating micro and macro perspectives, the analysis regards the agents in the economy as vertices, and the mutual relations among agents as edges, based on which large-scale complex networks are constructed. The existing network research in finance field is generally of two strains. The first one concentrates on clarifying the typological structure of the financial network (Iori et al., 2008；Cajueiro and Tabak, 2008；Boss et al., 2004). Due to data availability, most of the research modelled and analyzed the inter-bank market risk exposure and managed to figure out some typical complex network properties of financial networks. However, most of those research was still stuck in the statistical description of the network, with little concern about the impact of the typological properties on financial risk control. In addition, the majority of the networks only include one kind of vertices representing financial institutions. Bipartite networks including two kinds of vertices, such as banks and firms, were rarely constructed. For the second strain of the research, network analysis was used to explore the transmission path of financial risk and to identify the important financial entity with systemic significance (Upper, 2011；Hasman and Samartin, 2008). Many of those studies assumed that the financial network is a regular network or a random work, and some important characteristics of the actual financial work, such as scale-free and hierarchy properties, were ignored. Those ignorance would limit the external validity of the conclusions and the practical values to some extent. Besides, these studies mainly focused on the systematic significance of individual vertices, missing the impact of the network typological properties on systematic stability.

This study constructed a bipartite network based on actual financial transaction data to analyze the typological properties and transmission procedure for effective risk management. In this network, two kinds of vertices, financial institutions and listed firms, were included, which was rarely constructed in previous studies. In addition, the typological properties and risk transmission were examined in details. The time length of this study is fifteen years, which allowed us to evaluate the relationship between the evolution of network structure and the evolution of financial risk from a dynamic perspective. In addition, this study is based in China, the largest emerging country. It would add an important sample to the existing research, which was mainly developed countries based, for risk management in home and international market.

## 2. Data and Methodology

*2.1 The data*

This study used the credit data of listed firms in Shanghai and Shenzhen A-share markets from 2000 to 2014, which were drawn from CSMAR database, a leading financial database in China. The data was extracted with focus on the credit information between listed firms and financial institutions, including names of debtors and creditors, loan amount and currency. To ensure the validity and reliability of research, the original data was pre-processed: 1) excluding firms with no credit history; 2) excluding transactions without creditor or debtor information ; 3) excluding transactions without loan amount; 4) excluding listed firms under special treatment; 5) regularizing names of financial institutions. After the pre-treatment, we got 607 financial institutions, 1,777 listed firms and 23,133 loan records. Referring to the financial institutions classification standard by CBRC (China Bank Regulatory Commission), financial institutions were divided into 1) state-owned commercial banks; 2) policy banks; 3) nationwide joint-stock banks; 4) urban commercial bank and urban credit cooperatives; 5) rural cooperative bank and rural credit cooperatives; 6) foreign banks; 7) trust and financial firms. Among them, the number of trust and financial firms was the largest 246, 40.53% of the total institutions. The largest amount of loans were provided by large state-owned commercial banks, which was 2,152.176 billion yuan, 45.68% of the total amount. In accordance with the Guidance for Industry Classification of Listed Firms (2012 revised edition) issued by CSRC (China Securities Regulatory Commission), this study classified the industries of all the firms. The real estate development and operation industry has the largest number of the loan records and the largest amount of money, which was 1,996 and 80,308.68 million yuan respectively. For this industry, 8.63% of the loan records contributed to 11.56% loan amount, suggesting that important status of real estate industry in China. Figure 1 shows the changes in credit amount, number of financial institutions and firms in from 2000 to 2014. All the three indicators generally followed an increasing trend, while two twisted points could be seen in year 2006 and 2011.

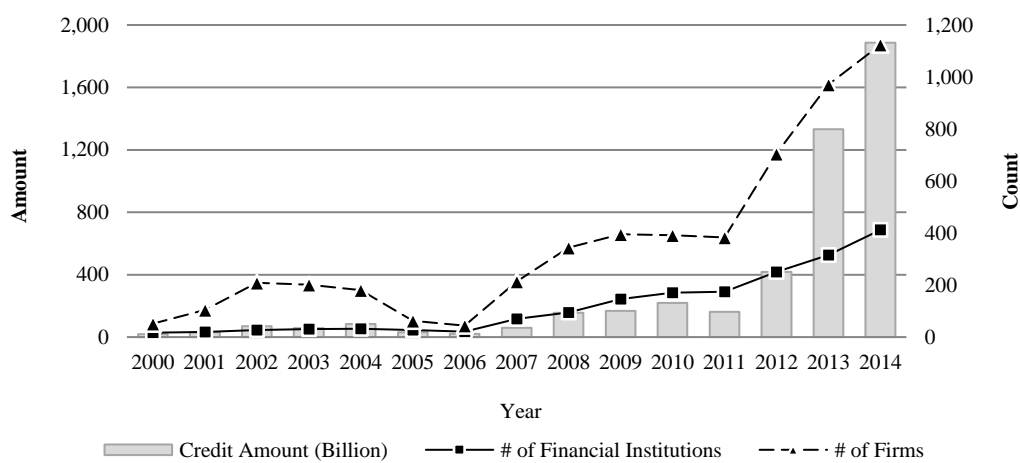

**Figure 1 The Number of Financial Institutions and Firms (2000-2014)**

*2.2 Construction of Financial Institution-Firm network and One-Mode Projections*

This study regards the credit market in China as a complex network. It is composed by a considerable number of financial institutions and firms as vertices, and connected by the credit relationships as edges. In this network, there are two kinds of vertices, financial institutions and firms, and the edges run only between vertices of unlike types, credit flow from financial institutions to firms. This type of network is called bipartite network or two-mode network, a special model in graph theory. Due to the special characteristics of bipartite networks, two one-mode projections can be created from the two-mode bipartite form. Because the information provided by the network based on two-mode data is quite different from the networks based on one-mode projections, both the two-mode and one-mode networks are usually analyzed jointly for better understanding (Borgatti and Halgin, 2011). In our study, the credit network based on two-mode data (financial institutions and firms) provides the information about the direction of credit flow across different industries or regions and the interdependent relationship between the financial institutions and firms, implying the possible contagion path and severity of financial crisis. For the projected credit sub-networks with one-mode data (financial institutions or firms) provides the information about the competition structure for high quality credit and risk-sharing mechanism among different financial institutions or firms, thus implying the hierarchical structure of the credit market and outlining those most crucial players. Both the bipartite network and the projected networks would be analyzed in this study.

Based on the credit data among financial institutions and firms, an undirected and unweighted network, and an weighted network were constructed respectively. As for the undirected and unweighted network, the adjacency matrix of t time period is the matrix with elements $a_{BF(i,j)}^t$ such that[①]:

$$a_{BF(i,j)}^t = \begin{cases} 1, & \text{if financial institution } i \text{ provided loan to firm } j \\ & \text{during } t \text{ time priod} \\ 0, & \text{Otherwise}; \end{cases}$$

For the weighted network, the adjacency matrix of t time period is the matrix with elements $w_{BF(i,j)}^t$ such that:

$$w_{BF(i,j)}^t = \begin{cases} > 0, & \text{the amount of loan provided by financial institution } i \\ & \text{to firm } j \text{ during } t \text{ time priod} \\ = 0, & \text{no loan provided by financial institution } i \\ & \text{to firm } j \text{ during } t \text{ time priod} \end{cases}$$

Based on the adjacency matrix and weighted adjacency matrix of financial institution-firm network and the characteristics of bipartite network, financial institution-financial institution network and firm-firm network could be projected. For financial institution-financial institution network, the adjacency matrix of time period t is the matrix with elements $a_{BB(i,j)}^t$ $(i \neq j)$ such that:

$$a_{BB(i,j)}^t = \begin{cases} 1, & \text{if financial institution } i \text{ and } j \text{ provided loan to at least} \\ & \text{one same company during } t \text{ time priod} \\ 0, & \text{Otherwise}; \end{cases}$$

---

[①] In all the matrixes in this study, financial institutions are denoted as B, while firms are denoted as F.

Similarly, for firm-firm network, the adjacency matrix of time period t is the matrix with elements $a_{FF(i,j)}^t$ $(i \neq j)$ such that:

$$a_{FF(i,j)}^t = \begin{cases} 1, & \text{if firm i and j received loan from at least one same} \\ & \text{financial institution during t time priod} \\ 0, & \text{Otherwise}; \end{cases}$$

In summary, for each time period, we can get two kinds of matrixes. One describes the credit relationship between financial institutions and firms, including an unweighted and a weighted matrix. The other one, projected from the bipartite, describes the relationship among financial institutions or firms. All these matrixes and the relationship among the entities provided basis for our empirical study.

### 3. Analysis of Typological Properties of Financial Institution-Firm Network

*3.1 Degree and Strength Distribution*

In complex network theory, vertex degree and strength describes the heterogeneity of vertex at individual level (Almaas et al., 2004). In financial institution-firm network, the degree of a financial institution vertex describes the number of firms it provided loans to. The degree of a firm vertex describes the number of financial firms it received loans from. The strength of a vertex describes the total amount of loans a financial institution provided or a firm received. The vertex degree and strength complements with each other to analyze how individual entities interact with each other from a micro perspective.

In t period, the degree and strength of financial institution vertices and firm vertices are defined as follows:

$$ND_{B_i}^t = \sum_J a_{BF(i,j)}^t ; \quad ND_{F_j}^t = \sum_I a_{BF(i,j)}^t \qquad (1)$$

$$NS_{B_i}^t = \sum_J w_{BF(i,j)}^t ; \quad NS_{F_i}^t = \sum_I w_{BF(i,j)}^t \qquad (2)$$

The time-evolution of degree and strength for financial institution-firm network conveyed that China has experienced a national credit expansion during the last fifteen years. The figure 2 shows that the average degree of each financial institution was volatile before 2006, and it was followed by a twisted upward. The average degree of each firm followed a steep increase from about 1.5 to a peak of 6.16. The figure 3 shows that vertex strength witnessed a similar trend. A sharp lifting after 2012 was observed for both financial institutions and firms

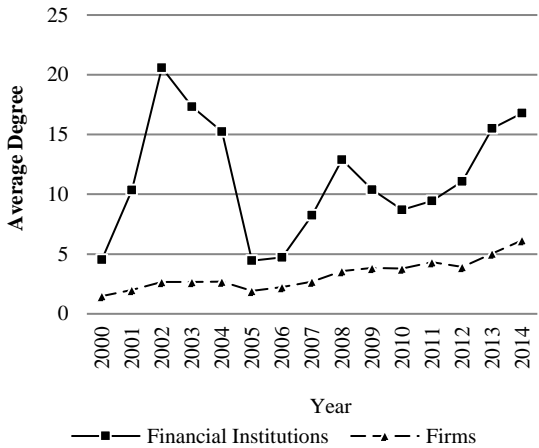

Figure 2 The Average Degree of Financial Institutions and Firms (2000-2014)

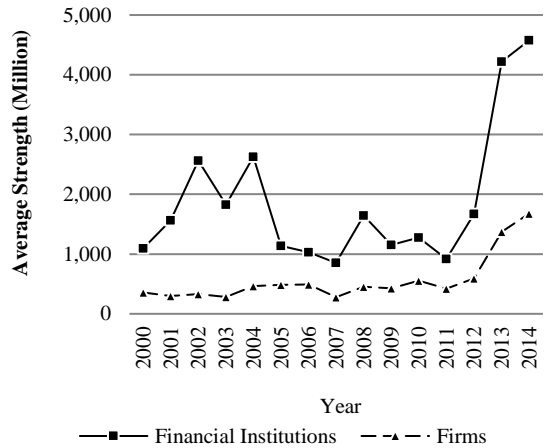

Figure 3 The Average Strength of Financial Institutions and Firms (2000-2014)

The semi-parameter estimation introduced by Clauset et al. (2014) was applied to analyze the degree and strength of vertices (see figure 4 and 5). The results showed that the distribution of degree and strength for both financial institutions and firms followed a power-law distribution, indicating that the financial institution-firm network was a typical scale-free network. Meanwhile, it was evident that the exponents of the fitted power-law distributions (degree and strength) of financial institutions displayed a lower level than those of firms.

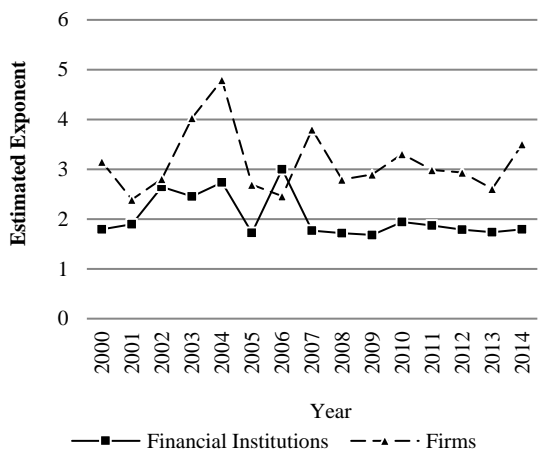

Figure 4 The Exponents of Fitted Power-law Distributions for Degrees Distribution (2000-2014)

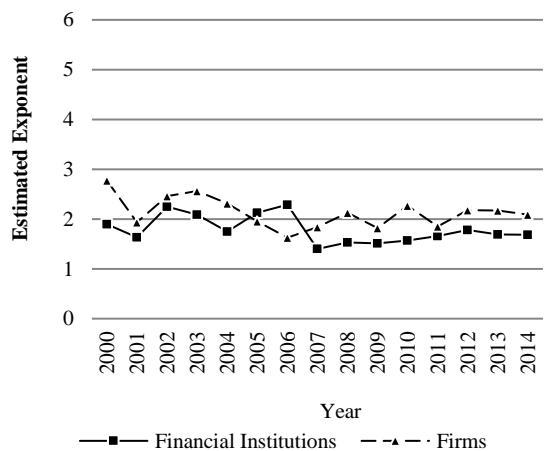

Figure 5 The Exponents of Fitted Power-law Distributions for Strengths Distribution (2000-2014)

In figure 6 and 7, we show the scaling of the strength versus the degree. In the case of financial institutions, the linear correlation coefficient between strengths and degrees is 0.8906 ($p<0.001$), while that for firms is 0.3183 ($p<0.001$). The correlation coefficient of financial institutions was higher than firms. It was mainly because financial institutions has much stronger need for risk diversification. To assure the safety of loans, financial institutions tend to spread large amounts of loans among many different firms. For the firms, the relationship between the amount of loans firms demand for and the number of financial institutions they ask to was positive but relatively weak. The reason is that the firms with large amount of loans prefer to borrow from many different resources (Ogawa et al., 2007). However, firms with a smaller amount of loans may also prefer multiple links. In addition, when firms borrow from many financial institutions, it does not mean that they must be seeking considerable amounts of loans. Instead, they may desire to have cooperation with different resources for better financing support in the future, which is especially true in China's financing environment.

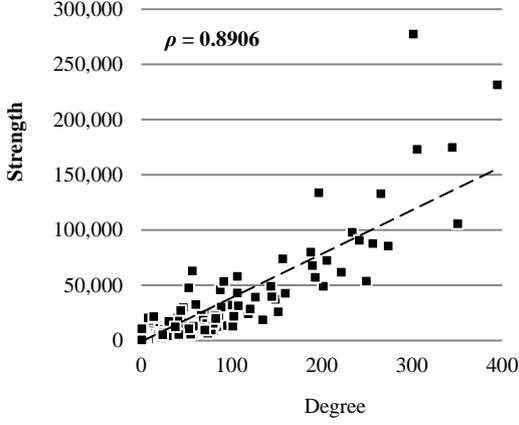
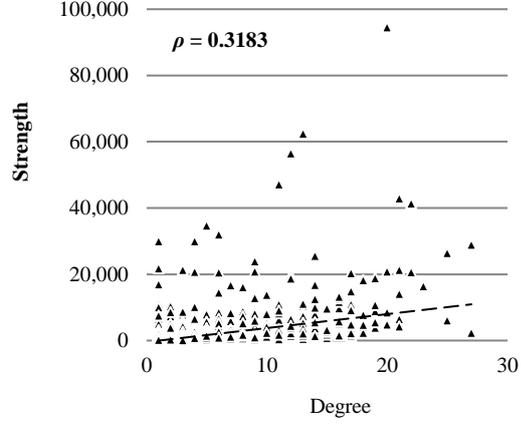

Figure 6  The Relationship between Degree and Strength of Financial Institutions (2000-2014)

Figure 7  The Relationship between Degree and Strength of Firms (2000-2014)

*3.2 Distribution of Relative Vertex Strength*

The relative vertex strength is introduced to measure to what extent the financial institutions and firms are interdependent with each other in the network. The dependence of firm *j* on financial institution *i* is defined as the share of credit amount it receives from *i* in its total liabilities. Similarly, the dependence of financial institution *i* on firm *j* is defined as the share of credit to *j* in its total lending. Hence, the relative strength increases with the relative importance of the financial institution as a creditor or firm as debtor. The relative strength for financial institution *i* and firm *j* t time period is defined respectively as:

$$RNS_{B_i}^t = \sum_J \left( w_{BF(i,j)}^t / \sum_I w_{BF(i,j)}^t \right) \qquad (3)$$

$$RNS_{F_i}^t = \sum_I \left( w_{BF(i,j)}^t / \sum_J w_{BF(i,j)}^t \right) \qquad (4)$$

In figure 8, we show that the relative strength of financial institutions fluctuated before 2006. It was followed by a remarkable plummeting from the peak of 0.3155 to 0.0891, while those of firms were stable at much lower level around 0.01. The analysis shows that the financial institutions generally held a relatively dominant positions in the credit relationships in China. Because of insufficient financial resources and risk aversion, small and medium sized firms were in relatively weak positions. In addition, due to the great influence of Chinese government policies, whether a firm can receive loans is also affected by policies, which also partly explain the low relative strength of firms. The table 1 shows that the relative strength of state-owned commercial banks was ranked first for 13 years out of 15 years, which suggested the prominent position in China's financial system. Policy banks were also ranked first in 2005 and 2006, which indicated the strong macro-control of Chinese government in this time period. For top ranked industries, manufacturing industries and energy-related industries were most frequently listed. From 2008 to 2014, cement manufacturing industries were ranked first for 4 times, which indicated the rapid development of infrastructure and construction. The Chinese economy was still largely driven by infrastructure development and real estate.

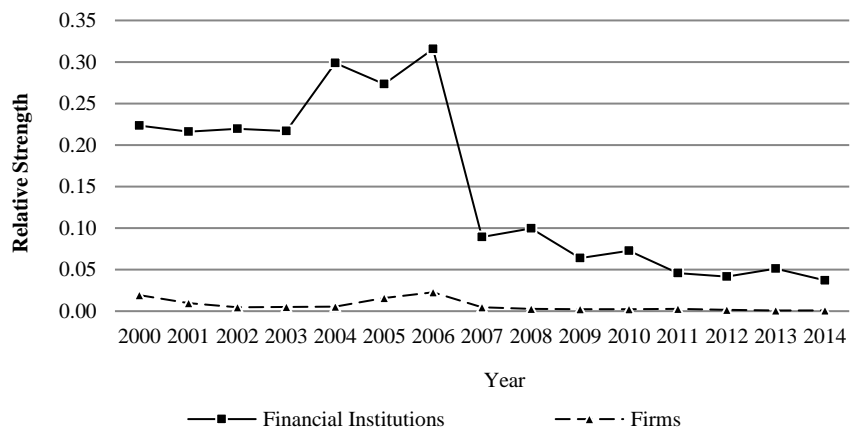

**Figure 8  The Relative Strength of Financial Institutions and Firms (2000-2014)**

**Table 1   The Top Ranking of Relative Strength for Financial Institutions and Firms (2000-2014)**

| Year | Type of Financial Institutions | RS | Industry of Firms | RS |
|---|---|---|---|---|
| 2000 | State-owned commercial banks | 0.6221 | Non-ferrous metal ore mining | 0.0588 |
| 2001 | State-owned commercial banks | 0.4954 | Transportation subsidiary service | 0.0302 |
| 2002 | State-owned commercial banks | 0.7215 | Information dissemination service | 0.0206 |
| 2003 | State-owned commercial banks | 1.0067 | Rubber manufacturing | 0.0578 |
| 2004 | State-owned commercial banks | 1.5008 | Metal products | 0.0401 |
| 2005 | Policy banks | 2.5240 | Wholesale of energy, materials, machinery and electronic equipment | 0.0370 |
| 2006 | Policy banks | 2.2656 | Medicine manufacturing | 0.0490 |
| 2007 | State-owned commercial banks | 0.5688 | Instrument, meter, stationery and office machine manufacturing | 0.0314 |
| 2008 | State-owned commercial banks | 1.1468 | Cement manufacturing | 0.0151 |
| 2009 | State-owned commercial banks | 0.9746 | Cement manufacturing | 0.0197 |
| 2010 | State-owned commercial banks | 0.6831 | Other electronic equipment manufacturing | 0.0088 |
| 2011 | State-owned commercial banks | 0.8843 | Cement manufacturing | 0.0103 |
| 2012 | State-owned commercial banks | 0.7463 | Cement manufacturing | 0.0116 |
| 2013 | State-owned commercial banks | 0.9541 | Coal mining and dressing | 0.0040 |
| 2014 | State-owned commercial banks | 0.9201 | Other public facilities services | 0.0049 |

*3.3   The Community Structure and Its Membership*

This section explores the regional network properties of the financial institution-firm network by conducting community detection. One network may be partitioned into different communities, with many edges connecting vertices in the same community and few connecting vertices among different ones (Fortunato, 2010). If the financial institution-firm network is completely connected, where any pair of financial institutions and firms has an equal chance to be linked, no conspicuous community structure would be expected. Otherwise, a remarkable number of communities may be detected. The community detection method introduced by Newman and Girvan (2004) to inspect the possible existence of communities in the network. Specifically, the modularity function of financial institution-firm network to be optimized is defined as:

$$\mathbb{Q} = \frac{1}{2M} \cdot \sum_{ij} [a_{BF(i,j)} - a^e_{BF(i,j)}] \cdot \delta(c_i, c_j) \quad (5)$$

Where $a_{BF(i,j)}$ is the corresponding element of its unweighted adjacency matrix, and $a^e_{BF(i,j)}$ is the probability of the presence of an edge between the t vertex $i$ and $j$ in the randomized null model. $M$ is the total number of existing edges, while the $\delta(c_i, c_j)$ is the indication function deciding the community membership: $\delta(c_i, c_j)$ equals 1 if the two vertex $i$ and $j$ are in the same community and 0 otherwise. The Newman-Girvan community detection algorithm shows a general picture of community structure at network-level. The next step of our study is to seek for information characterizing the obtained communities and their periodic evolution.

Table 2 listed the community detection results from 2000 to 2014. The number of the communities gradually rose from 10 to 36 with the two-year trough in 2003 and 2004, while the percentage of the largest community stabilized around 85% to 95% except for 69.57% in 2000. The fact that 89.98% of the network were connected together on average from 2000 to 2014 meant that the financial institution-firm network was a highly connected network. From the risk management prospective, such a community structure displayed the vulnerability to systematic financial risk. It indicated that the default risk burst from any given vertex could potentially spread among most of the vertices with unlimited cascades and feedback loop in a very short time (Haldane and May, 2011). Moreover, Table 2 summarized the most representative entity types, which had the largest amount of credits, in the largest community each year. It was noticed that the most representative type of the financial institutions was 'state-owned commercial banks' for the majority of the years. As for the firms, 'real estate development and operation' was identified as the most representative industry in nine out of fifteen years, and several manufacturing industries were ranked first from 2003 to 2008. In summary, while state-owned commercial banks played a dominant role in the largest community almost throughout the whole time period, the prominent industry changed from time to time. One interesting fact could be observed that the representative industry from 2008 to 2012 was real estate development and operation industry for four consecutive years, which indicated the large credit flows to this industry. It also provided a perspective to look at the highly disputed real estate industries in China.

Table 2  Summary of Information about the Largest Community (2000-2014)

| Year | Number of Communities | Percentage of the Largest Community | Representative Type of Financial Institution | Representative Industry of Firms |
|---|---|---|---|---|
| 2000 | 10 | 69.57% | State-owned commercial banks | Real estate development and operation |
| 2001 | 12 | 98.39% | State-owned commercial banks | Real estate development and operation |
| 2002 | 8 | 94.07% | State-owned commercial banks | Real estate development and operation |
| 2003 | 4 | 97.00% | State-owned commercial banks | Chemical material and products manufacturing |
| 2004 | 4 | 97.18% | State-owned | Electric power, steam and hot |

| Year | | | | |
|------|----|--------|--------------------------|----------------------------------------------|
| 2005 | 8  | 82.22% | Policy bank              | Daily use electronic equipment manufacturing |
| 2006 | 7  | 72.31% | State-owned commercial banks | Real estate development and operation |
| 2007 | 16 | 88.38% | State-owned commercial banks | Electric power, steam and hot water production and supply |
| 2008 | 14 | 93.62% | State-owned commercial banks | Daily use electronic equipment manufacturing |
| 2009 | 23 | 91.87% | State-owned commercial banks | Real estate development and operation |
| 2010 | 28 | 89.17% | Nationwide joint-stock bank | Real estate development and operation |
| 2011 | 24 | 91.40% | State-owned commercial banks | Real estate development and operation |
| 2012 | 28 | 93.93% | State-owned commercial banks | Real estate development and operation |
| 2013 | 27 | 95.57% | State-owned commercial banks | Civil engineering works construction |
| 2014 | 36 | 95.05% | State-owned commercial banks | Real estate development and operation |

## 4. The Analysis of the Topological Properties of Institution-Institution Network and Firm-Firm Network

From the analysis of topological properties of financial institution-firm network, we found that the financial institutions and firms were highly connected as indicated by the community structure, and asymmetrically connected, as indicated by the degree and strength distribution. It also implied that the local idiosyncratic shocks such as default risk were possible to proliferate through the whole economy and generate a sizable global disturbance. The projected credit sub-networks with only one-mode data provide the information from a supplementary perspective. As for the projected credit sub-networks based only on the financial institutions (institution-institution network), two connected institutions means that they offered credits to at least one common firm,. Both institutions jointly undertook the potential default risk of the common firm. For the sub-network based only on the firms (firm-firm network), two connected firms means that they received credits from at least one common financial institution. The two firms competed for credits. The topological properties analysis of the sub-networks shed some lights on the intensity and structure of the default risk sharing mechanism among different financial institutions, and competition among firms to obtain the credit. The clustering coefficient and the assortativity coefficient would be adopted for this analysis, where the former one provides insights on how dense the financial institutions were risk-shared or the firms competed with each other, and the later one indicated the interdependent mechanism.

The clustering coefficient measures the probability that two financial institutions or firms are connected with each other. The higher the clustering coefficient is, the more dense the sub-networks is. The clustering coefficient of is defined as:

$$CC_{\mathbb{G}} = \frac{\sum_i CC_{\mathbb{G}(i)}}{N} \qquad (6)$$

Where $CC_{\mathbb{G}(i)}$ is the individual clustering coefficient of vertex $i$[①], which is defined as:

$$CC_{\mathbb{G}(i)} = \frac{\#\{jk | k \neq j, j \in N_{\mathbb{G}(i)}, k \in N_{\mathbb{G}(i)}\}}{d_{\mathbb{G}(i)}(d_{\mathbb{G}(i)} - 1)/2} \qquad (7)$$

Here, $N$ is the number of vertex, and $d_{\mathbb{G}(i)}$ is the degree of vertex $i$. $d(v_i, v_j)$ equals to the shortest path length between vertex $i$ and $j$, and we further assume if vertex $i$ and $j$ are unconnected then $d(v_i, v_j)$ equals 0.

The assortativity coefficient measures the level of homophyly of the network, based on some vertex labeling or values assigned to vertices, such as degree. If the vertices show high tendencies to establish the link with other nodes that have similar (or dissimilar) scale of degrees as themselves, then the network is labeled as assortativity (or disassortativity) with a positive (or negative) assortativity coefficient (Newman, 2003). The assortativity coefficient is defined as:

$$r_{\mathbb{G}} = \frac{\frac{1}{|D_{\mathbb{G}}|} \cdot \sum k_i k_j - \left[\frac{1}{|D_{\mathbb{G}}|} \cdot \sum \frac{1}{2} \cdot (k_i + k_j)\right]^2}{\frac{1}{|D_{\mathbb{G}}|} \cdot \sum \frac{1}{2} \cdot (k_i^2 + k_j^2) - \left[\frac{1}{|D_{\mathbb{G}}|} \cdot \sum \frac{1}{2} \cdot (k_i + k_j)\right]^2} \qquad (8)$$

Where $|D_{\mathbb{G}}|$ is the number of existing edges in the network, and $k_i$ and $k_j$ are the degrees of vertex $i$ and $j$ respectively.

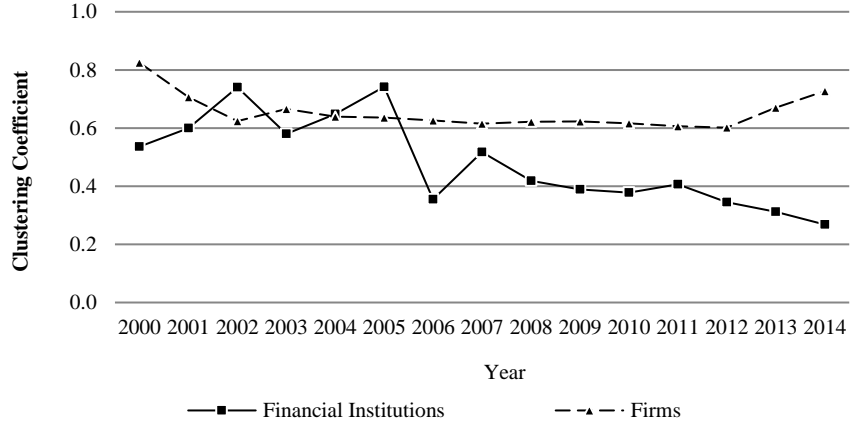

**Figure 9 The Clustering Coefficient of Two Projected Credit Subgraphs (2000-2014)**

---

[①] If the degree of vertex $i$ is no more than 1, the $CC_{\mathbb{G}(i)}$ is set to 0.

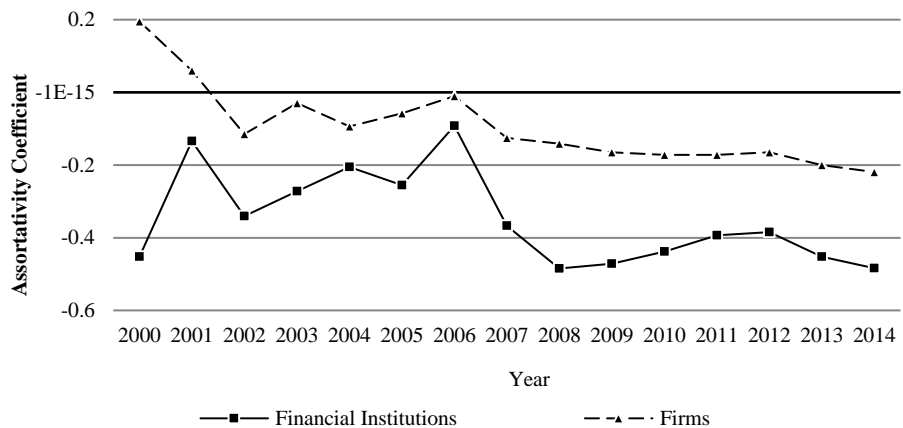

**Figure 10  The Assortativity Coefficient of Two Projected Credit Subgraphs (2000-2014)**

As for the institution-institution network, the clustering coefficient increased from 2001 to 2005, but it was followed by a decline since 2006. Meanwhile, the clustering coefficient of firm-firm network stabilized above 0.6 in the shape of an U. We interpret this as a clear signal of a long-standing credit market, where firms need to compete fiercely with other counterparties for a superior credit. The similar shaped behaviors were also observed in the assortativity coefficient. Figure 10 showed that the assortativity coefficients of the institution-institution network were all negative and experienced a rapid descent after 2006, while that of the firm-firm network were almost negative except for the first two years and fluctuated within a certain band between 0.2 and -0.1. These results suggested the disassortativity of both sub-networks. It implied that the firms tended to demand loans from large and small financial institutions simultaneously, and at the same time, the financial institutions preferred to cultivate the credit portfolio including the firms in various industries. A possible explanation is that the main players in the credit market before 2006 were the state-owned commercial banks, policy banks and nationwide joint-stock banks (called nationally-operated institutions), whose branches were distributed nationally, hence having more opportunities to offer the credit to one common firm. After 2006, the booming credit demands were hardly met only by those nationally-operated institutions. More and more urban commercial bank and urban credit cooperatives, rural cooperative bank and rural credit cooperatives, foreign banks and trust and financial firms (called locally-operated institutions) participated in the market. Due to the limited resources, and political and geographical constraints, the locally-operated institutions could only make the credit available to limited firms or in a limited area. Hence they were sparsely linked with the locally-operated institutions elsewhere and further stifled the excessive relationships. Figure 11 and 12 shows the change of the percentage of number and amount of credit provided by nationally-operated institutions from 2000-2014. The gradual downward trends were in supportive of the upsurge of the appearance of the locally-operated institutions in the credit market.

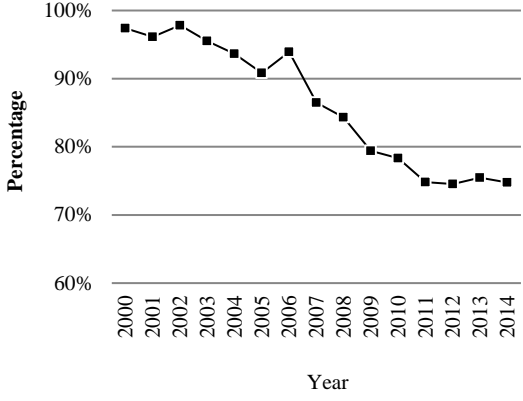
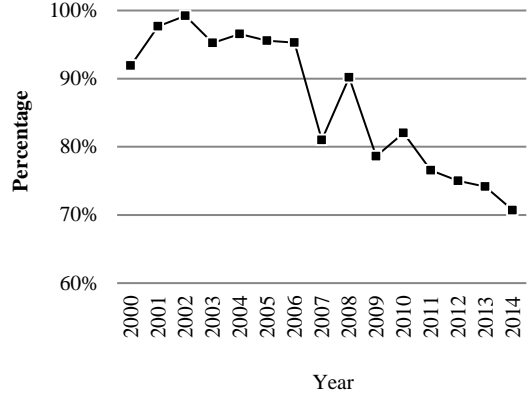

Figure 11  Percentage of Number of Credit Provided by Nationally-operated Institutions (2000-2014)

Figure 12  Percentage of Amount of Credit Provided by Nationally-operated Institutions (2000-2014)

## 5. Identification of Systematically Important Vertices in Credit Network

### 5.1 Construction of Credit Risk Score

An important purpose of analyzing the topological properties of credit network is to identify the entities of significantly systematic influence, which may potentially spread risks hastily to the whole system. Here we introduced feedback centrality (Galbiati et al., 2013) to construct credit risk score, CRS. The CRS measures the potential influence of an entity, either a financial institution or a firm, to others when a credit default occurs in this entity. The greater the score is, the more important this entity is for systematic stability. The entity with high CRS score are the key vertex for risk control.

First of all, the confronted risk of a financial institution and the confronted risk of a firm are defined as $\gamma_b$ and $\gamma_f$ respectively. $\gamma_b$ and $\gamma_f$ ranges from 0 meaning completely healthy assets, to 1 meaning bankruptcy. It is supposed that in t period, only the $i^{th}$ financial institution went bankrupt, all the other financial institutions and firms are healthy, i.e., $\gamma_{B_i}^t = 1$、$\gamma_{B_{j|j \neq i}}^t = 0$、$\gamma_{F_k}^t = 0$. In accordance with feedback centrality algorithm, the risk of all the entities in the t+1 period is as follows:

$$\gamma_{B_i}^{t+1} = 1 \qquad (9)$$

$$\gamma_{B_{j|j \neq i}}^{t+1} = \gamma_{B_{j|j \neq i}}^t + \sum_K w_{B_i F_k} \cdot \gamma_{F_k}^{t+1} \qquad (10)$$

$$\gamma_{F_k}^{t+1} = \gamma_{F_k}^t + \sum_J w_{F_k B_i} \cdot \gamma_{B_j}^{t+1} \qquad (11)$$

$w_{B_i F_k}$、$w_{F_k B_i}$ is defined as elements of risk diffusion matrix P, such that

$$w_{B_i F_k} = \frac{C_{B_i F_k}}{C_{B_i}}, \quad w_{F_k B_i} = \frac{C_{B_i F_k}}{C_{F_k}}$$

$C_{B_i F_k}$ measures the credit strength between the $i^{th}$ financial institution and the $k^{th}$ firm based on the credit amount. $C_{F_k}$ represents the total amount of loans the $k^{th}$ firm has received. Without loss of generality, we assume that there is nonexistence of multiple feedback loops. This is to say, when the risk of the $i^{th}$ financial institution is spread to the $k^{th}$ firm, and the risk is spread back to the $i^{th}$ financial firm according to formula (10), the risk would not be spread to the $k^{th}$ firm for the second time. Then, in accordance with formula (9), (10) and (11), when the $i^{th}$ financial institution went bankrupt in t period ($\gamma_{B_i}^t = 1$), the weighted risk index of all the financial institutions and firms in the network in t+1 period is as follows:

$$Risk_B = \frac{\sum_i C_{B_i} \cdot \gamma_{B_i}}{\sum_i C_{B_i}}, \quad Risk_F = \frac{\sum_k C_{F_k} \cdot \gamma_{F_k}}{\sum_k C_{F_k}}$$

The CRS (credit risk score) of the $i^{th}$ financial institution is defined as the sum of $Risk_B$ and $Risk_F$. Similarly, we got the CRS for the $k^{th}$ firm. Different from several widely-used algorithm such as Pagerank, HITS, the CRS involves the interactive mechanism among the two kinds of vertex in the bipartite network simultaneously. In addition, The initial shock ($\gamma_{B_i}^{t+1}$) to the whole system can be optimized according to the natures or conditions of the entities with high degree of flexibility, such that it could be set much higher in the situation of small-sized financial institutions than those of the large state-owned ones.

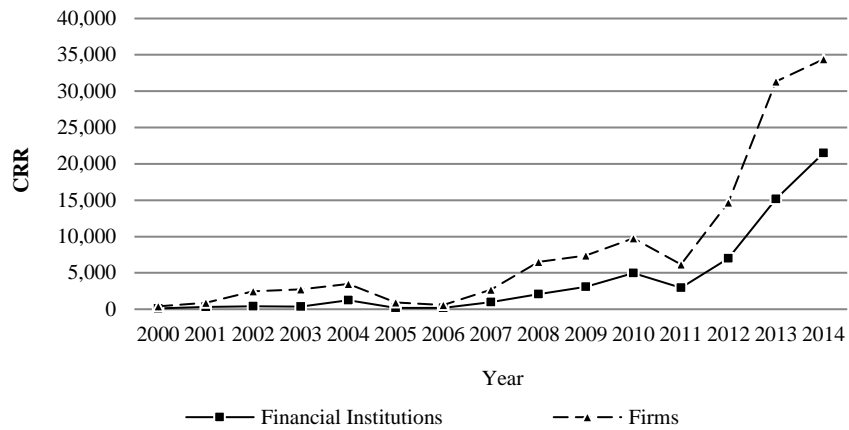

**Figure 13 The Trend of Credit Risk Score (CRS) (2000-2014)**

Figure 13 presents the simulation results of CRS for both financial institutions and firms from 2000 to 2014. From 2000 to 2014, the CRS generally followed an increasing trend. From 2000 to 2006, the CRS was less volatile and remained at a low level. From 2007 to 2010, the CRS gradually increased but still at contronable level. From 2011 to 2014, the CRS grew rapidly, and the average annual growth of fianancial institutions and firms reached 98.07% and 48.40% respectively. This swift increase may be largely due to the outbreak of the glocal economic crisis in 2008. In order to avoid further slowdown in China's economic growth rate, the Chinese government lauched massive economic and monetary stimulus packages. While the credit quota increased enormously, the qualification requirement by financial institutions of lending firms also decresed gradually, leading to a rapid growth of the overal CRS during this period.

**Table 3 The Top Ranking of CRS of Financial Institutions**

| Year | Financial Institutions | Type of Financial Institutions | CRS |
|---|---|---|---|
| 2000 | China Construction Bank | State-owned commercial bank | 54.31 |
| 2001 | Shanghai Pudong Development Bank | nationwide joint-stock bank | 58.00 |
| 2002 | China Construction Bank | State-owned commercial bank | 97.78 |
| 2003 | Shanghai Pudong Development Bank | nationwide joint-stock bank | 97.10 |
| 2004 | Industrial and Commercial Bank of China | State-owned commercial bank | 591.96 |

| Year | | | |
|---|---|---|---|
| 2005 | China Development Bank | Policy bank | 62.70 |
| 2006 | The Export-Import Bank of China | Policy bank | 45.42 |
| 2007 | China Everbright Bank | Nationwide joint-stock bank | 119.37 |
| 2008 | Hengfeng Bank | Nationwide joint-stock bank | 248.06 |
| 2009 | China Eastern Airlines Financial Service Company | Trust and financial firm | 278.50 |
| 2010 | Bank of Kunlun | Urban commercial bank and urban credit cooperatives | 553.62 |
| 2011 | The Export-Import Bank of China | Policy bank | 139.02 |
| 2012 | Changji Rural Commercial Bank | Rural cooperative bank and rural credit cooperatives | 292.98 |
| 2013 | China Resources Bank of Zhuhai | Urban commercial bank and urban credit cooperatives | 861.41 |
| 2014 | Tsingtao Brewery Finance Company | Trust and financial firm | 1410.35 |

Table 4  The Top Ranking of CRS of Firms

| Year | Firms | Industry of Firms | CRS |
|---|---|---|---|
| 2000 | Beijing Centergate Technologies Co., Ltd | Real estate development and operation | 160.00 |
| 2001 | Financial Street Holding Co., Ltd | Real estate development and operation | 132.93 |
| 2002 | Shanghai Fosun Pharmaceutical Co., Ltd. | Medicine manufacturing | 428.83 |
| 2003 | Beijing Gehua Catv Network Co., Ltd | Information dissemination service | 1043.48 |
| 2004 | Shanxi Zhangze Electric Power Co., Ltd | Electric power, steam and hot water production and supply | 1592.47 |
| 2005 | Tongfang Co., Ltd | Computer application service | 422.22 |
| 2006 | Zhongmin Energy Co., Ltd. | Paper making and paper products | 237.50 |
| 2007 | Huayu Automotive Systems Company Limited | Traffic equipment manufacturing | 269.20 |
| 2008 | Zhejiang China Commodities City Co., Ltd. | Comprehensive | 756.67 |
| 2009 | China Eastern Airlines Co., Ltd. | Air transport | 555.00 |
| 2010 | China Xd Electric Co., Ltd | Electric power, steam and hot water production and supply | 2381.62 |
| 2011 | Shanxi Lu'An Environmental Energy Development Co.,Ltd. | Coal mining and dressing | 788.00 |
| 2012 | Elion Clean Energy Company Limited | Chemical material and products manufacturing | 648.51 |
| 2013 | ZTE Corporation | Communications and related equipment manufacturing | 4005.28 |
| 2014 | Tsingtao Brewery Co., Ltd. | Beverage manufacturing | 2818.70 |

Table 3 and 4 listed the specific financial institutions and firms with the highest CRS in each year from 2000 to 2014. Firstly, the progressive increasing trend of the highest CRS values of financial institutions and firms both followed by an exponential growth model with the parameter equaling to 0.180 ($p<0.005$) and 0.158 ($p<0.005$). This implied that systematic risks were rapidly accumulated with the substantial expansion of credit since 2010. Secondly, the types of the financial institutions with the highest CRS and the industries of the firms with the highest CRS were much diverse than we expected, especially compared with the most representative ones in the largest communities as we discussed in section 3.3. The possible explanation is that CRS calculation takes account of the interactions of the vertex with other vertices in the network, while the most representative financial institution type or industry type was identified by amount of credits without considering interactions. For effective supervision of systematic risk, both the amount of credits and the interconnected nature of creditors and debtors from the network prospective should be considered. The community analysis and CRS analysis complement with each other to offer a comprehensive picture for effective systematic risk control.

Table 5  The Top Ranking of CRS by Types of Financial Institutions and Industries of Firms (2000-2014)

| Year | Type of Financial Institutions | CRS | Industry of Firms | CRS |
|---|---|---|---|---|
| 2000 | State-owned commercial banks | 13.71 | Real estate development and operation | 23.88 |
| 2001 | Urban commercial bank and urban credit cooperatives | 30.63 | Beverage manufacturing | 27.31 |
| 2002 | State-owned commercial banks | 33.16 | Electric power, steam and hot water production and supply | 38.30 |
| 2003 | State-owned commercial banks | 22.28 | Information dissemination service | 1043.48 |
| 2004 | State-owned commercial banks | 144.20 | Electric power, steam and hot water production and supply | 150.29 |
| 2005 | Policy bank | 33.90 | Computer application service | 211.15 |
| 2006 | Policy bank | 35.89 | Paper making and paper products | 130.75 |
| 2007 | Nationwide joint-stock banks | 27.10 | Traffic equipment manufacturing | 104.15 |
| 2008 | State-owned commercial banks | 75.69 | Broadcasting, movie and television | 102.64 |
| 2009 | Foreign bank | 35.06 | Air transport | 191.37 |
| 2010 | Nationwide joint-stock banks | 128.98 | Cement manufacturing | 1105.25 |
| 2011 | Policy bank | 73.36 | Cement manufacturing | 209.01 |
| 2012 | State-owned commercial banks | 53.89 | Cement manufacturing | 583.96 |
| 2013 | State-owned commercial banks | 210.83 | Cement manufacturing | 281.57 |
| 2014 | Policy bank | 158.43 | Leasing service | 398.27 |

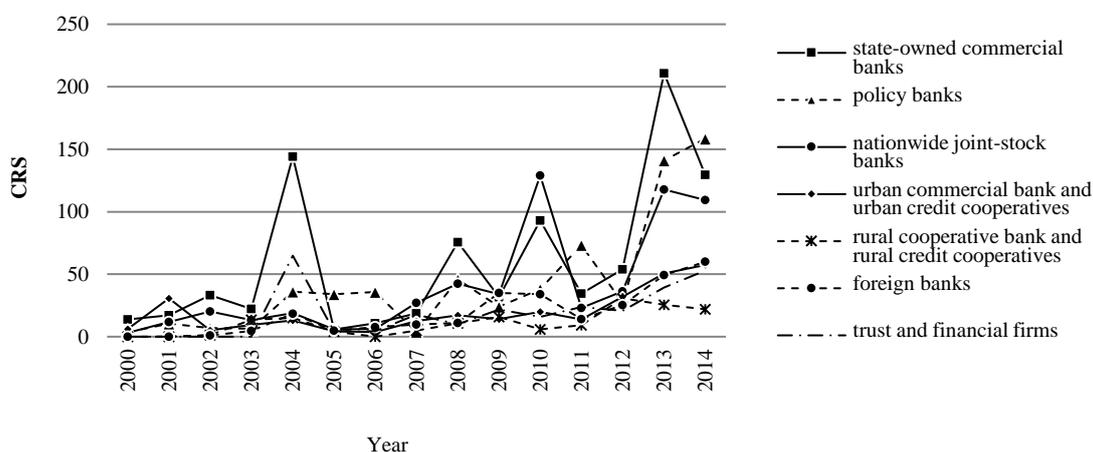

Figure 14  The Trend of Credit Risk Score (CRS) by Types of Financial Institutions (2000-2014)

Table 5 listed the type of financial institutions and the industry of firms with the highest average CRS in each year. For financial institutions, the ones with relatively high CRS were state-owned commercial banks, policy banks and nationwide joint-stock banks, which implied the crucial roles these banks played in the stability of China's financial system. For firms, the CRS of manufacturing industries were at high level, indicating that the manufacturing industries were still the pillar industries of the Chinese national economy. Figure 14 shows the CRS trend of financial institutions for 15 years. The CRS generally followed a rising trend. For state-owned commercial banks, policy banks and nationwide joint-stock banks, their CRS increased more significantly after 2007. However, for urban commercial bank and urban credit cooperatives, rural commercial bank and rural credit cooperatives, and trust and financial firms, the growth trend was relatively stable. In addition, the CRS of state-owned commercial banks was cyclical to some extent. This showed that, apart from the common functions of commercial banks, state-owned commercial banks also played a role in adjusting economic cycles for historical or political reasons, partly leading to high credit risk and large amount of non-performing assets for many years. On the whole, the average annual skewness of CRS was 3.71 and 9.094 for financial firms and firms respectively. The distribution of CRS was right skewed significantly. It means that the majority of entities had very limited influence on the systematic stability and only a small amount of financial firms would have substantial impact on the entire credit system if defaults.

It is also meaningful to investigate the effect of the topological properties of the vertices on the CRS, which shed the light on the determinants of the systematic risk at micro level from the network prospective. Besides the degree and strength mentioned in the previous sections, two network metrics are further introduced from a supplementary perspective. Firstly, in the networks, the greater the number of paths in which a vertex participates, the higher the importance of this vertex for the network. Thus, assuming that the interactions following the shortest paths between two vertices, the betweenness of the focal vertex is defined to quantify the importance of a vertex as (Costa et al., 2007):

$$Betweenness_k = \sum_{ij} \frac{\sigma(i,k,j)}{\sigma(i,j)}$$

Here, $\sigma(i,k,j)$ is the number of shortest paths between vertex $i$ and $j$ that pass through the focal vertex $k$. $\sigma(i,j)$ is the total number of shortest paths between $i$ and $j$, and the sum is over all pairs $i$, $j$ of distinct vertices.

Secondly, the average relative distance of each vertex from any other nodes in the network can be calculated as:

$$\bar{d}_i = N^{-1} \sum_{j=1}^{N} d_{ij}$$

By definition, the smaller the $\bar{d}_i$ is, the closer the vertex is to any other nodes, and the more center the vertex is located. The closeness of the focal vertex is defined as the reciprocal of this kind of relative distance as follows:

$$Closeness_i = N \cdot \left( \sum_{j=1}^{N} d_{ij} \right)^{-1}$$

In the context of financial institution and firm credit network, the betweenness measures the probability of the vertex serving as the bridge to spread the systematic risk, the closeness measures how fast the vertex could spread the systematic risk to other vertices in the network.

The analysis was conducted by estimating a fixed-effect panel specification while controlling time effect to explore the relationship between CRS and topological properties at vertex-level. The results (see Table 6 and Table 7) indicated that the degree and strength of the vertices exerted a positive impact on the CRS. It meant that the potential systematic risk originated from the vertex was associated with increasing access of the entities to the credit markets. Counterintuitively, the betweenness demonstrated an inverse relationship, implying the systematic risk was less likely to accumulate in the intermediate vertex during the risk proliferation. The possible reason was that the large amount of the shortest path through these bridge-like vertices could somehow disperse the embedded system risk to other vertices in the network. However, no significant relationship was found between the CRS and vertex-level closeness. Furthermore, the lagging effect of the topological properties measured by degree and strength was introduced to the analysis. The results revealed that only the strength of the vertex exerted first-order lagging effect to the CRS. Finally, it was to determine how periods of financial crisis moderate the relationship between topological properties of the vertices and their CRS. We refit the above model using the data after 2009[①]. The main conclusions drawn above were still unchanged, while the impact made by vertex degree was almost tripled escalated from 0.33912 to 0.9455. In addition, no significant lagging effect of were found. This might be caused by the dramatically-changed economic environments during the period captured by the time effect.

Table 6   The Relationships between CRS and Topological Properties of Vertex (Financial Institutions)

|  | Dependent: $CRR_{B_i}$ | | |
| --- | --- | --- | --- |
|  | (1) | (2) | (3) |
| Degree | 0.2709* | 0.3391*** | 0.9455** |
|  | (2.3252) | (3.5917) | (3.1814) |
| Strength | 0.00107*** | 0.00112*** | 0.00080** |
|  | (3.8231) | (3.6524) | (2.9102) |
| Betweenness | -0.0008** | -0.0008** | -0.0010** |
|  | (-2.886) | (-2.8143) | (-3.0256) |
| Closeness | -136970 | -80268 | -12324000* |
|  | (-0.5447) | (-0.2606) | (-2.1701) |
| Degree_Lag | - | 0.0561 | -0.64425 |
|  |  | (0.2341) | (-1.9146) |
| Strength_Lag | - | -0.00098** | -0.00071 |
|  |  | (-2.985) | (-1.9012) |
| Time Effect | Yes | Yes | Yes |
| $N$ | 1,810 | 1,012 | 754 |
| Adj. R-Squared | 0.1669 | 0.2029 | 0.301 |

Table 7   The Relationships between CRS and Topological Properties of Vertex (Firms)

|  | Dependent: $CRR_{F_i}$ | | |
| --- | --- | --- | --- |
|  | (1) | (2) | (3) |
| Degree | 2.1797*** | 2.9106*** | 2.9262*** |
|  | (9.7317) | (12.6211) | (10.5957) |
| Strength | 0.01003*** | 0.00496*** | 0.00514*** |
|  | (27.9718) | (14.069) | (12.8587) |

---

① The date of the four trillion yuan ($586 billion) stimulus package against the global financial crisis was unveiled at the later November in 2008. Therefore, we decided to use the data from 2009 to 2014 in order for data consistency.

| | | | |
|---|---|---|---|
| Betweenness | -0.00031* | -0.00035* | -0.00034* |
| | (-2.4717) | (-2.2422) | (-1.9692) |
| Closeness | -55211 | -158570 | -1633300 |
| | (-0.1749) | (-0.3125) | (-0.3655) |
| Degree_Lag | - | -0.1375 | -0.08416 |
| | | (-0.507) | (-0.2616) |
| Strength_Lag | - | 0.00105*** | 0.00103*** |
| | | (3.8812) | (3.4205) |
| Time Effect | Yes | Yes | Yes |
| N | 5,383 | 2,668 | 1,998 |
| Adj. R-Squared | 0.2271 | 0.3374 | 0.3453 |

*5.2 Immune Effectiveness Analysis of the CRS*

The financial risk regulatory mechanism based on Basel agreement usually regards the capital requirements as the core indicator of the financial sector regulation. However, the above CRS analysis suggests that each entity in the credit system exerts significantly different influence on systematic stability. Therefore, it is not sufficient to supervise all the financial institutions with only one type of indicators, such as reserve and capital adequacy ratio, and with the same standard. The financial risk transmission mechanism in the credit network should not be ignored. One effective way proposed by this study is to calculate the CRS of each entity in the credit market, and to take stringent regulatory measures towards the entities with high CRS, which would allocate regulatory resources more efficiently.

This section draws on the transmission theory of infectious diseases. Through the comparison of influences exerted by different types of risks on system stability, it is to verify the effectiveness of the suggested CRS method. In complex network, the network stability refers to the network connectivity after deletion of certain vertices and edges (Albert et al., 2000). In our study, we measures the system stability by deleting certain vertices, which simulates that certain entities defaults and go bankrupt after some shocks in the credit system. Four indexes were selected to measure the connectivity of the network: 1) Scale of the largest connected subgraph, SLCS; 2) Number of Communities, NC; 3) Graph Density, GD; 4) Average Path Length，APL. At the initial stage, the network connectivity was at the best status, with the large SLCS, large GD, few NC and short APL. While the network was attacked continuously, the network was dispersed into subsets. The network connectivity got worse, with smaller SLCS and GD and larger NC and APL.

Referring to the method used by Crucitti et al. (2013), this study used selective attack and random attack to delete vertices. For the former method, vertices would be deleted according to the risk conditions, from vertices with high CRS to vertices with low CRS. If the CRS is good measurement of a vertex's systematic risk, the system stability would be damaged to the largest extent in the case of selective attack.

Table 8  The Change of Network Connectivity under Two Attack Strategies

| | SLCS | NC | GD | APL |
|---|---|---|---|---|
| | \multicolumn{4}{c}{Financial Institutions} | | | |
| CRS | 4.080% | -68.552% | 3.576% | -7.440% |
| Random | 1.871% | -0.110% | 1.713% | -2.525% |

|  | Firms | | | |
|---|---|---|---|---|
| CRS | 0.466% | -0.185% | 0.173% | -0.239% |
| Random | 0.192% | -0.001% | 0.066% | 0.036% |

As it can be seen from table 8[①], compared with random attack, the network was damaged to the largest extent in the case of selective attack. In this scenario, both SLCS and GD decreased fast, and NC and APL increased fast. The analysis suggested that the Chinese credit system had relatively high stability in the case of random attacks, while it had relatively low stability in the case of selective attacks. It also implied that the CRS is good measurement of an entity's systematic risk in the credit network.

## 6. Conclusions and Policy Suggestions

This paper investigates China's credit market by drawing the credit data between listed firms in Shanghai and Shenzhen A-share markets and financial institutions for 15 years from CSMAR database. By viewing the China's credit market as an interdependent bipartite network on actual financial transaction data, the financial institution-firm network and its projected sub-networks were constructed using the theory of complex network, whose topological properties were analyzed in details. In addition, credit risk score (CRS), a network-based measurement, was introduced to measure the potential systematic risk embedded in a financial institution or a firm, providing new insights on the risk control from a network perspective.

Firstly, we measured some topological properties of the financial institution-firm network such as degree, strength and relative strength distribution, and the community structure and its membership. We found that the financial institution-firm network was a typical scale-free network, and both degree and strength featuring significant heterogeneity witnessed a gradual upward due to the national credit expansion during the last fifteen years. From the analysis of community structure, it was found that the state-owned commercial banks played a dominant role in the credit market almost throughout the whole time period, while the prominent industry changed from time to time. The financial institutions and firms were highly connected as captured by the community structure and they were asymmetrically connected, as captured by the degree and strength distribution. It implied that the local idiosyncratic shocks such as default risk were possible to proliferate through the whole economy and generate a sizable global disturbance.

Next, we focused on the topological properties of the two projected credit sub-networks with only one-mode data (financial institution or firm), finding that both the institution-institution and firm-firm network demonstrated the characteristics of disassortativity. It suggested that the temporal distribution of credit was remarkable asymmetric and was still dominated by large state-owned financial institutions, leaving a long-standing credit market with firms highly competitive for financial resources.

Finally, the credit risk score (CRS) was introduced by simulation to identify the systematically important vertices in terms of systematic risk control. The increasing CRS at the network and vertex level implies that systematic risks were rapidly accumulated with the substantial expansion of credit, especially after the launch of massive economic and monetary stimulus packages since

---

① The benchmark of the changes in table 8 was the measurement of network connectivity before the attack.

2009. Meanwhile, the relationship between a vertex's typological properties and its CRS was analyzed. It indicated that the more access an entity had to the credit market, the higher CRS the vertex had. The more bridge-like an vertex was, which was measured by betweenness, it was more likely that it could disperse the embedded systematic risk to other vertices, and therefore it had smaller CRS. At last, by comparing the stability of the credit network under the different attack strategies, the effectiveness of CRS was verified.

Policy suggestions were proposed in accordance with the above findings. As we discussed in section 5, the systematically important vertices with higher CRS should be paid close attention in the macro-prudential supervision. Based on this idea, two regulatory strategies are recommended. One is to set the minimal capital requirement on the entity. The reserve capital of the financial institutions and firms with high credit risk score should be no less than a critical threshold of their credit exposures:

$$\bar{c}_{it} = max(c_{i(t-1)}, \theta_{it} L_{i(t-1)})$$

Where $\bar{c}_{it}$ and $c_{i(t-1)}$ are the reserve capital of $i^{th}$ entity at the t and t-1 period, and $L_{i(t-1)}$ is its credit exposures at t-1 period. The rates of provisions for reserve capital ($\theta_{it}$) need to be timely adjusted by supervisions to ensure that the reserve capital of $i^{th}$ entity meet the requirements.

The second strategy is to set the minimum capital-risk exposure ratio. In accordance with the strength of edges in the credit network, for each financial institution and firm, set the reserve capital requirement based on its largest weighted credit. More specifically,

$$\bar{c}_{it} = max(c_{i(t-1)}, max_{j|i \neq i}(E_{ij(t-1)})/\theta_{it})$$

Where $E_{ij(t-1)}$ is the strength of edge based on credit amount between entity i and entity j. The above two regulatory strategies adopted the perspectives from vertices and edges, taking account of the impact of heterogeneity on financial risk transmission path and scope.

The second suggestion is to leverage the modularity of the credit network for better stability of the credit system. By cutting the credit network into independent sub networks (hierarchy modules), it would cut down the risk transmission path, helping decrease the diffusion speed of systematic risks. This suggestion is consistent with the 'ring-fence principle' proposed in the Volcker Rule adopted in U.S. in 2014, and with similar accords in the UK and Europe, such as Vickers Report. The ring-fence separates the business of commercial banks from other businesses, preventing the spread of high risk business to traditional business. Differentiate the entities with different impacts on financial stability and expose different supervision standard accordingly. Network modeling and analysis would be of good use for this distinction. It is also important to reduce complexity and increase transparency of financial systems to guard against systemic and regional financial risks.

The last suggestion is to optimize the market allocation of credit resources to better serve the real economy. This suggestions is particularly for emerging countries. The financial service sector needs to be more competitive and inclusive. In boosting financial market competition and easing market access, the development of small financial institutions should be supported to enrich competitive offers of financial resources. In addition, the compensation mechanism of credit risk for small and micro firms and financial institution-firm cooperation platform should be established to motivate financial institutions to serve small and micro firms, resolving their financial difficulties. Besides, formulate different supervision standards for rural financial risks. Under controllable risk, policy-oriented financial resources should be guided to increase the support the development of rural areas, with reference to other countries' prior experience.